\documentclass[aps,pra,reprint,twocolumn,floatfix,groupedaddress,superscriptaddress]{revtex4-2}
\usepackage{pstricks,graphicx,amsmath,bbm,mathrsfs,amssymb,psfrag,pifont,times,mathptmx}
\usepackage[utf8]{inputenc}
\usepackage{hyperref}
\usepackage[english]{babel}
\usepackage{color}
\usepackage{ulem}
\usepackage{siunitx}
\usepackage{mathtools}
\usepackage{comment} 
\def\Id{{\mathbbm 1}}

\def\erf{{\rm erf}}

\def\PSQL{P_{\rm SQL}}
\def\PHel{P_{\rm H}}
\def\PK{P_{\rm K}}
\def\PHYNORE{P_{\rm HY}}
\def\PDISP{P_{\rm disp}}
\def\PHYB{P_{\rm hyb}}

\def\p{{\rm p}}
\def\disp{{\rm disp}}
\def\hyb{{\rm hyb}}

\usepackage{xparse}
\NewDocumentCommand{\ceil}{s O{} m}{%
  \IfBooleanTF{#1} 
    {\left\lceil#3\right\rceil} 
    {#2\lceil#3#2\rceil} 
}

\begin{document}
\title{Beating the standard quantum limit for binary phase-shift-keying discrimination with a realistic hybrid feed-forward receiver}
\author{Michele N.~Notarnicola}
\affiliation{Dipartimento di Fisica ``Aldo Pontremoli'',
Universit\`a degli Studi di Milano, I-20133 Milano, Italy}
\affiliation{INFN, Sezione di Milano, I-20133 Milano, Italy}

\author{Stefano~Olivares}
\email{stefano.olivares@fisica.unimi.it}
\affiliation{Dipartimento di Fisica ``Aldo Pontremoli'',
Universit\`a degli Studi di Milano, I-20133 Milano, Italy}
\affiliation{INFN, Sezione di Milano, I-20133 Milano, Italy}

\date{\today}
\begin{abstract}
We propose a hybrid feed-forward receiver (HFFRE) for the discrimination of binary phase-shift-keyed coherent states based on the appropriate combination of the displacement feed-forward receiver (DFFRE) and a homodyne-like setup employing a low-intensity local oscillator and photon-number-resolving detectors. We investigate the performance of the proposed scheme addressing also realistic scenarios in the presence of non-unit quantum detection efficiency, dark counts and a visibility reduction. The present HFFRE
outperforms the DFFRE
in all conditions, beating the standard quantum limit in particular regimes.
\end{abstract}
\maketitle

\section{Introduction}
The design of optimized receivers for discrimination of non orthogonal quantum states is a central task for both quantum communications \cite{Helstrom1970, Proakis2001, Cariolaro2015, Bergou2010} and continuous-variable quantum key distribution \cite{Grosshans2002, Gisin2002, Leverrier2009, Denys2021, Notarnicola2022, Notarnicola2023-LD, Notarnicola2023-KB, Jarzyna2023}. In particular, in the framework of optical communications, coherent-state discrimination plays a central role, as encoding based on these states attains the classical capacity of a lossy bosonic channel \cite{Giovannetti2004}. Nevertheless, their non orthogonality \cite{Olivares2021} makes it challenging to design an optimum receiver achieving the minimum error probability. 

In practical situations, conventional receivers employing either homodyne or heterodyne measurements \cite{Olivares2021} achieve the so-called shot-noise limit or standard quantum limit (SQL), that is the best error probability achievable by semi-classical means \cite{Helstrom1970, Proakis2001, Cariolaro2015}. Nevertheless, Helstrom proved the existence of an optimum receiver, beating the SQL and reaching the ultimate precision limit \cite{Helstrom1976, Cariolaro2015}. Accordingly, Helstrom's theory identifies the minimum error probability compatible with quantum mechanics laws, namely, the Helstrom bound.

For binary phase-shift-keying (BPSK), this optimum receiver may be implemented into practice by the Dolinar receiver \cite{Dolinar1973, Lau2006, Assalini2011}, based on continuous-time measurements, optical feedback and time-dependent displacement operations. Although theoretically optimal, the practical implementation of the Dolinar receiver exhibits several drawbacks. Indeed, feedback control requires fast detectors and electronics, with response times much lower than the symbol repetition rate. Moreover, the visibility reduction associated with an imperfect displacement operations crucially affects the performance, reducing significantly any quantum advantage \cite{Lau2006}.

As a consequence, the task is to design suboptimal receivers being more feasible with the current technology. At first, Kennedy proposed a near-optimum receiver employing a ``nulling" displacement operation followed by on-off detection \cite{Kennedy1973}, reaching twice the Helstrom bound. An improved version of this scheme has been developed by Takeoka and Sasaki by optimizing the magnitude of the displacement operation, reducing further the error probability in the low-energy regime \cite{Takeoka2008}. 
More recently, there has been designed a hybrid near-optimum receiver (HYNORE) performing a conditional ``nulling" displacement, whose phase is chosen by exploiting the a priori information obtained by performing homodyne-like detection on a fraction of the incoming signal \cite{Notarnicola2023}. The HYNORE is near optimum and beats the Kennedy receiver in the high-energy regime, reducing the gap with respect to the Helstrom bound.
Further improvements have also been obtained by Sych and Leuchs with the displacement feed-forward receiver (DFFRE) based on the splitting of coherent states \cite{Sych2016}. Here the incoming signal is split into $N$ copies thanks to an array of beam splitters, thereafter the displacement-photon counting scheme employed in \cite{Takeoka2008} is implemented on each copy, optimizing the displacement amplitude via feed-forward Bayesian inference. This approach allows to reach lower error probabilities in the low-energy regime, but converges towards the Kennedy receiver for large energies. Remarkably, this scheme approximates the Dolinar receiver in the limit $N \gg 1$ of infinite copies \cite{Assalini2011}.

In this paper we show that the HYNORE scheme can be successfully used to sensibly improved the performance of the DFFRE proposed in Ref.~\cite{Sych2016} . After dividing the incoming signal at a beam splitter with optimized transmissivity, we perform homodyne-like detection on the reflected branch and implement the feed-forward on the transmitted one as in \cite{Sych2016}, but now exploiting the information from the homodyne-like to further enhance the Bayesian inference method. The resulting error probability is closer to the Helstrom bound in the low-energy regime and then saturates to the HYNORE error probability, outperforming the original  receiver \cite{Sych2016} for all energies.
When realistic inefficiencies of the detectors, namely, quantum efficiency, dark counts and visibility reduction, are considered, the enhanced receiver is still able to beat the SQL in particular regimes, maintaining some quantum advantage, although the Helstrom bound is not reachable anymore.

The structure of the paper is the following. In Sec.~\ref{sec: Binary} we recall the main features of quantum discrimination theory. Then, in Sec.~\ref{sec: NearOptRec} we briefly outline the two relevant receivers above mentioned, namely, the HYNORE \cite{Notarnicola2023} and the 
DFFRE \cite{Sych2016}. Sec.~\ref{sec: Hyb} presents our proposal of the hybrid feed-forward receiver (HFFRE) and discusses its performance with respect to the displacement feed-forward one. Instead, Sec.~\ref{sec: Ineff} is devoted to the analysis of the proposed receiver in the presence of realistic inefficiencies. Finally, in Sec.~\ref{sec: Concl} we summarize the obtained results and draw the conclusions.

\section{Basics of coherent-state discrimination}\label{sec: Binary}

In this paper we address BPSK discrimination, namely, discrimination of the two coherent states
\begin{align} 
|\alpha_k\rangle = \big|e^{i (k+1) \pi} \, \alpha\big\rangle \, , \qquad (k=0,1) \, ,
\end{align} 
$\alpha>0$, having the same energy $\alpha^2$ but phase-shifted by $\pi$ \cite{Cariolaro2015, Bergou2010, Helstrom1976}. 
The two states $|\alpha_0\rangle=|-\alpha\rangle$ and $|\alpha_1\rangle=|\alpha\rangle$ are generated with equal a priori probabilities $\pi_0=\pi_1=1/2$. The task is to implement a receiver, that is a binary positive-operator-valued measurement (POVM) $\{\Pi_0, \Pi_1\}$ associated with a decision rule, such that outcome ``0" infers state $|\alpha_0\rangle$ and outcome ``1" infers $|\alpha_1\rangle$. Because of the nonzero overlap between the encoded states, the final decision may be incorrect and, thus, any receiver exhibits an error probability.

Conventional receivers in optical communications employ homodyne detection, whose associated error probability is the SQL:
\begin{align}\label{eq:SQL}
\PSQL= \frac12 \bigg[1-\erf(\sqrt{2}\alpha) \bigg] \, ,
\end{align}
$\erf(x)$ being the error function.

On the contrary, in ideal conditions the Dolinar receiver \cite{Dolinar1973} is the optimum receiver, reaching the minimum error probability allowed by quantum mechanics, namely, the Helstrom bound \cite{Helstrom1976, Cariolaro2015}:
\begin{align}\label{eq:Helstrom}
\PHel &= \frac12 \Bigg( 1- \sqrt{1-4\pi_0\pi_1 \big|\langle \alpha_0|\alpha_1\rangle\big|^2}\Bigg) \notag \\
&= \frac12 \Bigg( 1- \sqrt{1-e^{-4\alpha^2} }\Bigg) \, .
\end{align}

Due to the nontrivial implementation of this receiver, many suboptimal schemes have been proposed.
Among them, the most relevant benchmark is provided by the Kennedy receiver \cite{Kennedy1973}. It consists in the application of a fixed displacement operation $D(\alpha)$, mapping states 
\begin{align}
|\alpha_0\rangle \rightarrow |0\rangle \quad \mbox{and} \quad |\alpha_1\rangle \rightarrow |2\alpha\rangle  \, .
\end{align}
This is the so-called ``nulling" displacement, since one of the two signals is displaced into the vacuum state.
Thereafter, on-off detection is performed on the output signals and the decision criterion reads: ``off"~$\rightarrow$~``0" and ``on"~$\rightarrow$~``1". Thus, an error occurs when an ``off" result is retrieved from state $|\alpha_1\rangle$, leading to the error probability
\begin{align}\label{eq:Kennedy}
\PK =  \frac12 \big| \langle 0 | 2 \alpha \rangle \big|^2 = \frac{e^{-4\alpha^2}}{2} \, .
\end{align}
The Kennedy receiver is said to be near-optimum, as $\PK \approx 2 \PHel$ for $\alpha^2 \gg1$.
An improved version of the receiver may be obtained by optimizing the displacement amplitude \cite{Takeoka2008}, i.e. substituting the ``nulling" displacement $D(\alpha)$ with a generic $D(\beta)$, $\beta>0$, whose value is optimized to minimize the overall error probability.

Further examples of near-optimum receivers will be presented in the next section.

\section{Near-optimum receivers}\label{sec: NearOptRec}
In this section we present in more detail the two receivers employed to design our hybrid feed-forward receiver, namely, the HYNORE and the DFFRE. For a better clarity, the performance analysis of these receivers will be performed in the following section together with the hybrid feed-forward one.
\subsection{Hybrid near-optimum receiver (HYNORE)}\label{sec: HYNORE}
\begin{figure}[t]
\includegraphics[width=0.95\columnwidth]{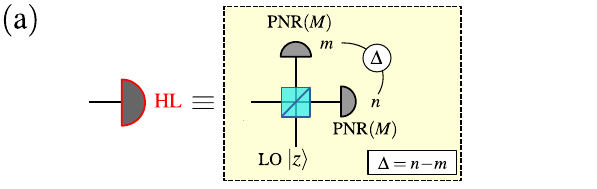}\\[5ex]
\includegraphics[width=0.95\columnwidth]{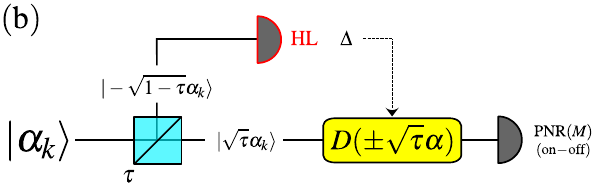}
\caption{(a) Implementation of homodyne-like detection. The signal is mixed at a balanced beam splitter with a low-intensity local oscillator (LO); thereafter PNR($M$) detection is performed on both branches. (b) Scheme of the HYNORE. We split the incoming signal at a beam splitter with transmissivity $\tau$, we perform homodyne-like detection on the reflected fraction, whose outcome conditions the displacement operation on
the transmitted part.}\label{fig:01-Hynore}
\end{figure}

An example of a near-optimum receiver beating the Kennedy is the HYNORE, we recently proposed in \cite{Notarnicola2023}. This scheme employs a homodyne-like (HL), or weak-field homodyne detection, that is a homodyne setup where the usual proportional photodetectors generating macroscopic photocurrents (such as p-i-n photodiodes, namely conventional analog photodetectors composed of an appropriate junction of semiconductors with different doping) are replaced with photon-number-resolving (PNR) detectors with finite resolution and, consequently, a low local oscillator (LO) is considered \cite{Allevi2017, Bina2017, Donati2014}. HL detection has also been implemented experimentally on different platforms, such as hybrid photodetectors \cite{Olivares2019}, transition-edge sensors \cite{Thekkadath2020, Nehra2020} and silicon photomultipliers \cite{Chesi2019:1, Chesi2019:2}.

The scheme is depicted in  Fig.~\ref{fig:01-Hynore}(a). We start mixing the incoming signal at a balanced beam splitter with the LO excited in the coherent state $|z\rangle$, $z>0$, then we perform PNR detection on both the output branches, obtaining outcomes $n$ and $m$, respectively, and, finally, we compute the difference photocurrent $\Delta=n-m$.
However, realistic PNR detectors have a finite resolution $M$, i.e. they can only resolve up to $M$ photons. To highlight this feature in the following they are referred to as PNR$(M)$ detectors. We describe PNR$(M)$ detection by the $M$-valued POVM $\{\Pi_0, \Pi_1, \ldots, \Pi_M \}$, with:
\begin{align}
\Pi_n = 
\left\{\begin{array}{ll} 
 | n \rangle\langle n| &\mbox{if}~ n=0,\ldots, M-1 \, , \\[2ex]
 \displaystyle \Id - \sum_{j=0}^{M-1} | j \rangle\langle j| &\mbox{if}~ n=M .
\end{array}
\right.
\end{align}
Accordingly, one has $-M \le \Delta \le M$.

Given an input coherent state $|\zeta\rangle$, $\zeta \in \mathbb{C}$,
following the previous outline, the HL probability distribution reads
\begin{align}\label{eq:Skell}
\mathcal{S}_\Delta&(\zeta) =
     \sum_{n,m=0}^{M} p_n\big(\mu_{+}(\zeta)\big) \
    p_m\big(\mu_{-}(\zeta)\big) \, \delta_{(n-m),\Delta}
\end{align}
where $\delta_{k,j}$ is the Kronecker delta, 
\begin{align}\label{eq:mupm}
\mu_{\pm}(\zeta)= \frac{|\zeta \pm z|^2}{2}\, ,
\end{align}
is the mean energy on the two output branches, respectively, and
\begin{align}\label{eq:PnPNRM}
    p_n(\mu) =
    \left\{\begin{array}{l l}
    {\displaystyle e^{-\mu} \ \frac{\mu^{n}}{n!}}  & \mbox{if}~n<M \ , \\[2ex]
    {\displaystyle 1- e^{-\mu} \sum_{j=0}^{M-1} \frac{\mu^{j}}{j!}} & \mbox{if}~n = M \ ,
    \end{array}
    \right.
\end{align}
being the probability of obtaining the outcome $n$ from PNR$(M)$ detection.
We note that Eq.~(\ref{eq:PnPNRM}) represents a truncated Poisson distribution; consequently, in the limit $M\gg1$ Eq.~(\ref{eq:Skell}) approaches the Skellam distribution \cite{Allevi2017, Bina2017}.

The HL scheme described above may be exploited to design the HYNORE, as depicted in Fig.~\ref{fig:01-Hynore}(b).
HYNORE performs HL measurement on a fraction of the input signal to gain some information on the phase of the optical field which is, then, used to improve the performance of a Kennedy or a displacement receiver \cite{Notarnicola2023}. The amount of the signal portion is optimized to minimize the overall error probability together with the amplitude of the LO of the HL. More in details, the incoming signal $|\alpha_k\rangle$, $k=0,1$, is split at a beam splitter with transmissivity~$\tau$. At first, we perform HL detection on the reflected branch $|\alpha^{(r)}_k\rangle=|-\sqrt{1-\tau} \alpha_k\rangle$, obtaining the outcome $\Delta$. Then, according to the value of $\Delta$ we decide the sign of a displacement operation $D(\pm \sqrt{\tau} \alpha)$ to be performed on the transmitted signal $|\sqrt{\tau} \alpha_k\rangle$. If $\Delta \ge 0$ it is more likely that $|\alpha_0\rangle$ was sent, therefore we choose $D(\sqrt{\tau} \alpha)$, otherwise we apply $D(-\sqrt{\tau} \alpha)$. Finally, we perform on-off detection, which may be still realized via PNR$(M)$ detection. The final decision rule is depicted in Table~\ref{tab:01-DecRule}.
Eventually, the overall error probability writes:
\begin{align}\label{eq:Phynore}
\PHYNORE = \min_{\tau, z} \, \PHYNORE(\tau,z)\, ,
\end{align}
where
\begin{align}
\PHYNORE(\tau,z) =& \frac{e^{-4 \tau \alpha^2}}{2} \,\Bigg[
\sum_{\Delta=-M}^{-1} {\cal S}_\Delta\big(\alpha^{(r)}_0\big)
+ \sum_{\Delta=0}^{M} {\cal S}_\Delta \big(\alpha^{(r)}_1\big) \Bigg] \, ,
\end{align}
and the optimization over the transmissivity $\tau$ and the LO amplitude $z$ of the HL scheme can be carried out.
\begin{table}[tb]
\centering
\begin{tabular}{c |c}
    outcomes & decision \\  \hline
    $\Delta \ge 0$ $\quad$ off \, & ``0" \\ 
    $\Delta < 0$ $\quad$ on \, & ``0" \\ 
    $\Delta < 0$ $\quad$ off \, & ``1" \\ 
    $\Delta \ge 0$ $\quad$ on \, & ``1" \\ \hline
\end{tabular}
\caption{Decision strategy for the HYNORE depicted in Fig.~\ref{fig:01-Hynore}(b).}\label{tab:01-DecRule}
\end{table}

As discussed in \cite{Notarnicola2023}, the best performance of the HYNORE is obtained in the presence of a high-intensity local oscillator (homodyne limit) and full photon number resolution. However, from a practical point of view this would require to employ different types of detectors for the two components of the setup: two proportional photodiodes producing macroscopic photocurrents to implement the standard homodyne measurement on the reflected signal, and a PNR detector for the displacement receiver on the transmitted branch. Moreover, pulsed homodyne detection is preferable for an experimental realization at telecom wavelength, due to the reduced response time of the measurement \cite{Raymer1995, Hansen2001, Zavatta2002}.
On the contrary, employing HL and low-intensity local oscillator provides a more fascinating solution since a near-optimum receiver is obtained with the use of sole PNR detectors.

\subsection{Displacement feed-forward receiver (DFFRE)}\label{sec: Disp}
\begin{figure}[t]
\includegraphics[width=0.95\columnwidth]{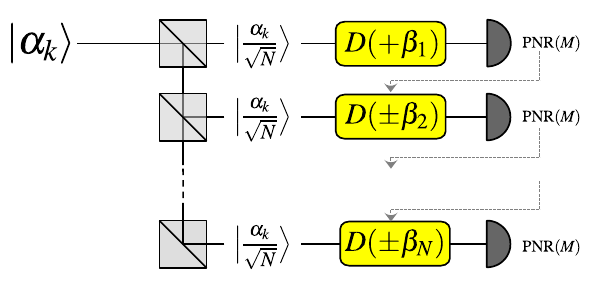}
\caption{Scheme of the DFFRE proposed in \cite{Sych2016}. The incoming signal $|\alpha_k\rangle$, $k=0,1$, is split into $N$ copies and undergoes a sequence of conditional displacements followed by photon counting. The first copy undergoes a positive displacement, whereas the sign of the subsequent displacements is decided via Bayesian inference.}\label{fig:02-DispScheme}
\end{figure}

The second near-optimum receiver, namely the DFFRE proposed in \cite{Sych2016}, is depicted in Fig.~\ref{fig:02-DispScheme} and exploits the splitting of the input state $|\alpha_k\rangle$ into $N$ modes, or (rescaled) copies, that is
\begin{align}
|\alpha_k\rangle \rightarrow \bigotimes_{j=1}^{N} |\alpha_k^{(j)}\rangle \, ,
\end{align}
where $|\alpha_k^{(j)}\rangle=|\alpha_k/\sqrt{N}\rangle$. Then, each copy undergoes an optimized conditional  displacement followed by PNR($M$) detection.
We start by displacing the first copy $|\alpha_k^{(1)}\rangle$ by $D(\beta_1)$, with amplitude $\beta_1>0$ maximizing the correct decision probability, thereafter we perform PNR$(M)$ detection on the output signal $|\alpha_k/\sqrt{N}+\beta_1\rangle$. According to the maximum a posteriori probability (MAP) criterion based on Bayesian inference, the PNR$(M)$ measurement outcome $n$ is used to choose the sign of the optimized conditional displacement to be performed on the second copy. In other words, we infer the state ``0" or ``1" associated with the maximum a posteriori probability given the outcome $n$ \cite{Notarnicola2023, Sych2016, DiMario2019}. If ``0" is inferred we displace the second copy $|\alpha_k^{(2)}\rangle$ by $D(\beta_2)$, otherwise we apply $D(-\beta_2)$, where $\beta_2>0$ is chosen to maximize the correct decision probability, too. Then, we perform again photodetection and repeat the process until the $N$-th copy.

With ideal detectors, the previous criterion is equivalent to performing on-off detection on each displaced copy. The $j$-th copy, $j=1,\ldots, N$, undergoes the displacement operation $D(\sigma_j \beta_j)$, where $\beta_j>0$ is the optimized amplitude and $\sigma_j=\pm 1$ is the sign of the displacement. The first displacement has a fixed sign, namely, $\sigma_1=+1$. The other values of $\sigma_j$ are assigned according to the following decision rule: if we get outcome ``off" from the $(j-1)$-th measurement we set $\sigma_j= \sigma_{j-1}$, otherwise if a ``on" is retrieved we switch $\sigma_j= -\sigma_{j-1}$. Ultimately, the outcome obtained from the last copy determines the final decision. Therefore, the outcome ``off" infers state $|-\sigma_{N} \alpha\rangle$, outcome ``on" infers state $|\sigma_{N} \alpha\rangle$, $\sigma_{N}$ being the sign of the last displacement. 
A complete explanation of the functioning of this receiver is performed in Appendix~\ref{app:derivation}, together with a detailed derivation of the associated error probability.

The discrimination error probability of the displacement feed-forward receiver depends on the number of copies $N$ and reads:
\begin{align}\label{eq:Pdisp}
\PDISP^{(N)} = 1-{\cal P}_{\disp}^{(N)} \,,
\end{align}
${\cal P}_{\disp}^{(j)}$, $j=1,\ldots,N$, being the probability of performing a correct decision after $j$ steps (see Appendix~\ref{app:derivation}), namely:
\begin{align}\label{eq:PcorrDISP}
{\cal P}_{\disp}^{(j)} &= 
\max_{\beta_{j}} \Bigg\{ {\cal P}_{\disp}^{(j-1)}
\, q_{\rm off}\Big( \lambda_{-}^{(j)}(\alpha)\Big) \notag \\
&\hspace{2cm}  
+\left[1-{\cal P}_{\disp}^{(j-1)}\right] q_{\rm on}\Big( \lambda_{+}^{(j)}(\alpha)\Big) 
\Bigg\} \, ,
\end{align}
where
\begin{align}\label{eq:q01}
q_{\rm off}(x) = e^{-x} \quad \mbox{and} \quad q_{\rm on}(x) = 1-e^{-x} \,,
\end{align}
are the probabilities of ``off" and ``on" results, respectively, and
\begin{align}\label{eq:lambdapm}
\lambda_{\pm}^{(j)}(\alpha)= \Big|\beta_j \pm\frac{\alpha}{\sqrt{N}}\Big|^2 \, ,
\end{align}
is the mean photon number of the resulting displaced copies.
As reported in Eq.~(\ref{eq:PcorrDISP}), we remark that the value of the displacement amplitude $\beta_j$, $j=1,\ldots,N$, is chosen to maximize the correct decision probability at each step $j$ of the feed-forward scheme.

\section{Hybrid feed-forward receiver (HFFRE)}\label{sec: Hyb}

\begin{figure}[t]
\includegraphics[width=0.99\columnwidth]{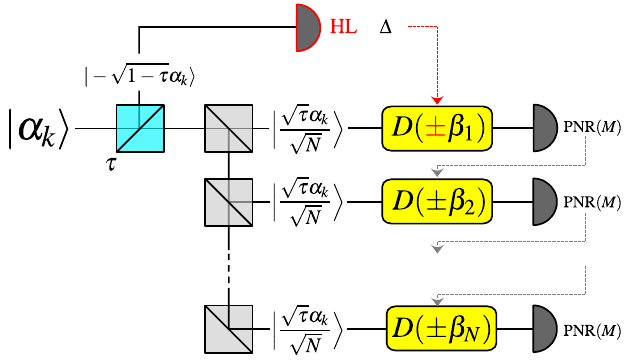}
\caption{Scheme of the HFFRE. We split the incoming signal $|\alpha_k\rangle$, $k=0,1$, at a beam splitter of variable transmissivity $\tau$. We perform HL detection on the reflected branch, whereas we implement the displacement feed-forward setup on the transmitted one. We exploit the HL outcome to decide the sign of the displacement operation on the first copy of the transmitted signal.}\label{fig:03-HybFF}
\end{figure}

The two setups discussed in Sec.~\ref{sec: NearOptRec} can be suitably merged to construct a hybrid feed-forward receiver, the HFFRE, as depicted in Fig.~\ref{fig:03-HybFF}. The insight is to exploit a HL measurement to guide the choice of the first displacement operation sign in the displacement feed-forward receiver.
Thus, we divide the incoming signal $|\alpha_k\rangle$, $k=0,1$, at a beam splitter with variable transmissivity $\tau$, such that
\begin{align}
|\alpha_k\rangle \rightarrow |\alpha_k^{(r)}\rangle \otimes |\alpha_k^{(t)}\rangle = |-\sqrt{1-\tau}\alpha_k\rangle \otimes |\sqrt{\tau}\alpha_k\rangle \, .
\end{align}
The reflected signal $|\alpha_k^{(r)}\rangle$ undergoes HL detection with outcome $\Delta$. Then, we split the transmitted state $|\alpha_k^{(t)}\rangle$ into $N$ copies, $|\alpha_k^{(t)}/\sqrt{N}\rangle$, and implement the same procedure described in Sec.~\ref{sec: Disp}. The only difference with respect to the displacement feed-forward receiver lies in the displacement operation performed on the first copy. Indeed, the difference photocurrent $\Delta$ provides us with a priori information exploitable to decide the sign of the first optimized displacement operation, according to the following rule:
\begin{align}
\left\{\begin{array}{ll}
\Delta \ge 0 \quad &\rightarrow \quad \mbox{apply } D(\beta_1) \, \\[1ex]
\Delta < 0 \quad &\rightarrow \quad \mbox{apply }  D(-\beta_1) \, ,
\end{array}
\right.
\end{align}
$\beta_1>0$. Displacements on the other copies are still conditioned on the outcomes of the $(j-1)$-th PNR($M$) measurement.

Given the previous considerations, the probability of performing a correct decision ${\cal P}_{\rm hyb}^{(j)}(\tau,z)$ after $j$ steps gets the same form of Eq.~(\ref{eq:Pdisp}):
\begin{align}\label{eq:PcorrHYB}
{\cal P}_{\rm hyb}^{(j)}&(\tau,z) = \notag\\[1ex]
&\max_{\beta_{j}}
\Bigg\{ {\cal P}_{\rm hyb}^{(j-1)}(\tau,z)  q_{\rm off}\Big( \lambda_{-}^{(j)}(\sqrt{\tau}\alpha)\Big)
\notag\\[1ex]
& \hspace{0.5cm}+ \left[
1 -  {\cal P}_{\rm hyb}^{(j-1)}(\tau,z)
\right]q_{\rm on}\Big( \lambda_{+}^{(j)}(\sqrt{\tau}\alpha)\Big) \Bigg\} \, ,
\end{align}
albeit with a different initial condition, that is:
$${\cal P}_{\rm hyb}^{(0)}(\tau,z)=\frac12 \Bigg[ \sum_{\Delta=-M}^{-1} {\cal S}_\Delta\Big(\alpha^{(r)}_1\Big)
+ \sum_{\Delta=0}^{M} {\cal S}_\Delta\Big(\alpha^{(r)}_0\Big) \Bigg] \, , $$ 
corresponding to the probability of correct decision after the HL measurement. Clearly, if $\tau=1$ we retrieve the results of the DFFRE.

As both $\tau$ and $z$ are free parameters, after $N$ copies the error probability reads
\begin{align}\label{eq:Phyb}
\PHYB^{(N)} = 1- \max_{\tau,z}{\cal P}_{\rm hyb}^{(N)}(\tau,z)  \, .
\end{align}

\begin{figure}[t]
\includegraphics[width=0.95\columnwidth]{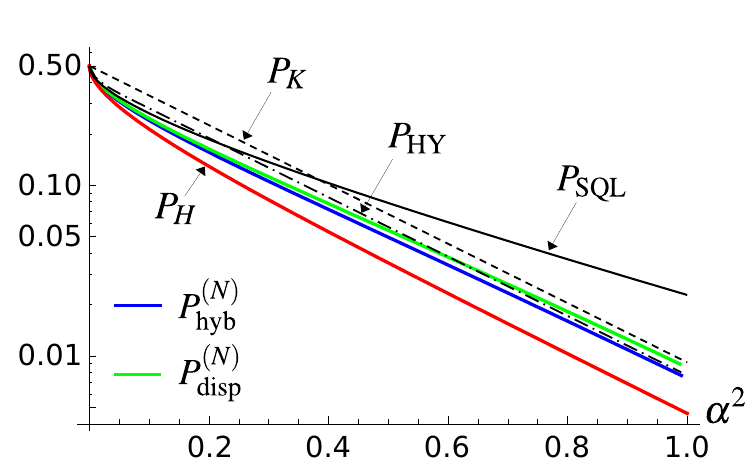}
\caption{Log plot of $\PHYB^{(N)}$ and $\PDISP^{(N)}$ as a function of the signal energy $\alpha^2$ for $N=1$. The PNR resolution is $M=2$. $\PSQL$, $\PHel$, $\PK$ and $\PHYNORE$ refer to the SQL~(\ref{eq:SQL}), the Helstrom bound~(\ref{eq:Helstrom}), and the error probabilities of the Kennedy receiver~(\ref{eq:Kennedy}) and the HYNORE~(\ref{eq:Phynore}), respectively.}\label{fig:04-ErrorP}
\end{figure}

Plots of $\PHYB^{(N)}$ and $\PDISP^{(N)}$ are depicted in Fig.~\ref{fig:04-ErrorP} as a function of the input energy $\alpha^2$. The HFFRE outperforms the DFFRE,
$\PHYB^{(N)} \le \PDISP^{(N)}$. Both the receivers are near-optimum and beat the SQL for all energies, but we have different asymptotic scalings. Indeed, for $\alpha^2 \gg 1$, the DFFRE approaches the Kennedy receiver, $\PDISP^{(N)} \approx \PK$, whereas the HFFRE reaches the HYNORE, $\PHYB^{(N)} \approx \PHYNORE$. As a consequence, exploiting information on both the phase and the photon statistics of the field proves to be a powerful tool to reduce the error probability.

\begin{figure}[t]
\includegraphics[width=0.95\columnwidth]{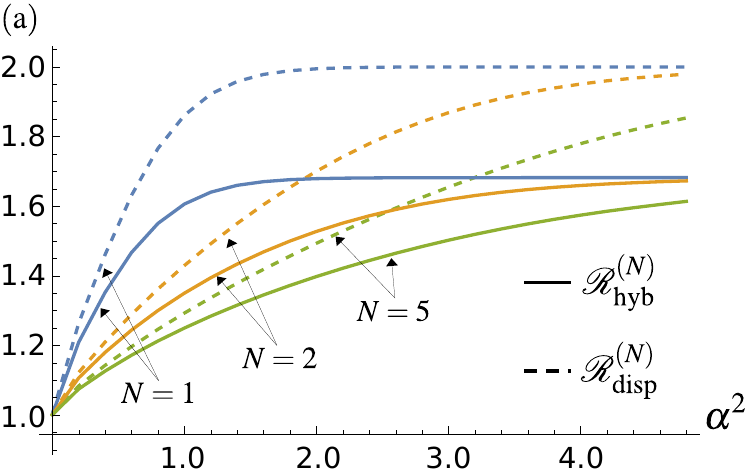}\\[2ex]
\includegraphics[width=0.95\columnwidth]{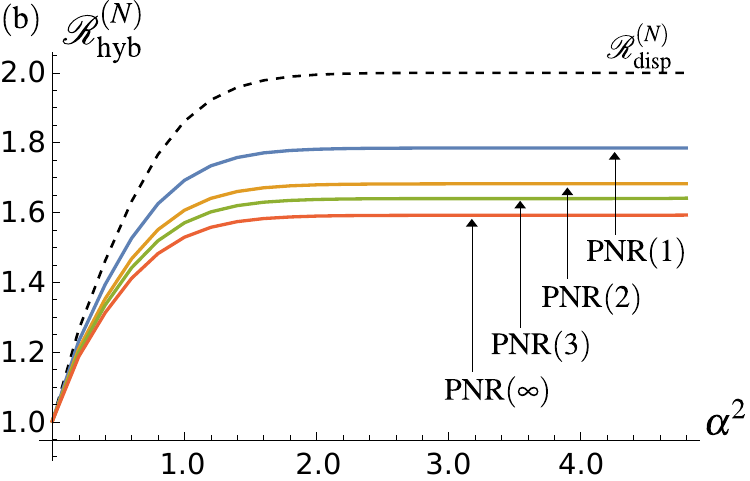}\\
\caption{(a) Plot of ${\cal R}_{\p}^{(N)}$, $\p=\disp,\hyb$, as a function of the signal energy $\alpha^2$ for different number of copies $N$. The PNR resolution is $M=2$. (b) Plot of ${\cal R}_{\hyb}^{(N)}$ as a function of the $\alpha^2$ for $N=1$ and different PNR resolutions $M$. The dashed line corresponds to ${\cal R}_{\disp}^{(N)}$ for $N=1$.}\label{fig:05-RatioHEL}
\end{figure}

Furthermore, by increasing the number of copies $N$ the performance of both the feed-forward receivers improves for $\alpha^2 \ll 1$, coming closer to the Helstrom bound~(\ref{eq:Helstrom}), as emerges by computing the ratio
\begin{align}
{\cal R}_{\p}^{(N)} = \frac{P_{\p}^{(N)}}{\PHel} \, , \quad (\p=\disp,\hyb) \, ,
\end{align}
plotted in Fig.~\ref{fig:05-RatioHEL}(a).

In the regime $\alpha^2 \ll 1$, the larger the number of copies, the smaller the ratio ${\cal R}_{\p}^{(N)}$, whereas in the asymptotic limit $\alpha^2 \gg 1$ the displacement and hybrid receiver converge to Kennedy and HYNORE, respectively, regardless the value of $N$. Moreover, the ratio for the hybrid receiver ${\cal R}_{\hyb}^{(N)}$ may be further reduced by increasing the PNR resolution $M$, as shown in Fig.~\ref{fig:05-RatioHEL}(b). In particular, the asymptotic ratio is reduced for greater values of $M$ and reaches its minimum value for PNR$(\infty)$ detectors, i.e. ideal photodetectors, in which case the HL distribution in Eq.~(\ref{eq:Skell}) becomes a Skellam distribution.

\section{Analysis of detection inefficiencies}\label{sec: Ineff}
We now consider a more realistic scenario by discussing how the typical imperfections in PNR detection affect the performance of the proposed hybrid receiver. In particular, we consider a non-unit quantum efficiency $\eta\le1$ of the PNR($M$) detectors, as well as the presence of dark counts. Moreover, since the displacement
operation is realized into practice by letting the signal interfere with a suitable LO at a beam splitter, we also address the effects of non-unit visibility $\xi\le1$.

As one may expect, in these conditions neither the DFFRE nor the HFFRE are able to approach the Helstrom bound anymore. Accordingly, a new goal emerges, that is to show whether or not these receivers are still able to beat the SQL~(\ref{eq:SQL}) even in the presence of realistic imperfections. Indeed, in this case we would get a robust quantum advantage with respect to the best receiver achievable with semi-classical means.

For the sake of simplicity, in the following we will perform the analysis by considering the sole hybrid receiver. In fact, the error probability associated with the displacement one may be retrieved in an analogous way by setting $\tau=1$.
 
\subsection{Quantum efficiency $\boldsymbol\eta$}\label{sec:QEff}

As coherent states are considered as inputs, the presence of a quantum efficiency $\eta \le 1$ requires only to rescale the coherent amplitudes of all the measured pulses by a factor  $\sqrt{\eta}$, as no mixedness is introduced at the detectors.
Thereafter, in the hybrid scheme of Fig.~\ref{fig:03-HybFF} the HL probability distribution of the reflected signal $|\alpha^{(r)}_k\rangle$ becomes ${\cal S}_{\Delta} \Big(\eta \alpha^{(r)}_k\Big)$
with the $\mu_{\pm}$ in Eq.~(\ref{eq:mupm}).
The effect is the same on the transmitted branch,  where the average photon numbers of the displaced copies $\lambda_{\pm}$, see Eq.~(\ref{eq:lambdapm}), are replaced by $\eta \lambda_{\pm}$. In turn, the correct decision probability ${\cal P}_{\rm hyb}^{(j)}(\eta;\tau,z)$ becomes
\begin{align}
&{\cal P}_{\rm hyb}^{(j)}(\eta;\tau,z) =\notag\\[1ex]
&\hspace{0.5cm}
\max_{\beta_{j}} \Bigg\{ {\cal P}_{\rm hyb}^{(j-1)}(\eta;\tau,z)\,
q_{\rm off}\Big( \eta \lambda_{-}^{(j)}(\sqrt{\tau}\alpha)\Big)\notag \\[1ex]
&\hspace{1cm}
+\bigg[1 - {\cal P}_{\rm hyb}^{(j-1)}(\eta;\tau,z)\bigg] q_{\rm on}\big( \eta\lambda_{+}^{(j)}(\sqrt{\tau}\alpha)\big) \Bigg\} \, ,
\end{align}
to be solved with the initial condition
$${\cal P}_{\rm hyb}^{(0)}(\eta;\tau,z)=
\frac12 \Bigg[ \sum_{\Delta=-M}^{-1} {\cal S}_\Delta\Big(\eta\alpha^{(r)}_1\Big)
+ \sum_{\Delta=0}^{M} {\cal S}_\Delta\Big(\eta\alpha^{(r)}_0\Big) \Bigg] \, , $$ 
and the associated error probability reads
\begin{align}\label{eq:Phyb_eta}
\PHYB^{(N)}(\eta) = 1- \max_{\tau,z}{\cal P}_{\rm hyb}^{(N)}(\eta;\tau,z)  \, .
\end{align}

\begin{figure}[t]
\includegraphics[width=0.95\columnwidth]{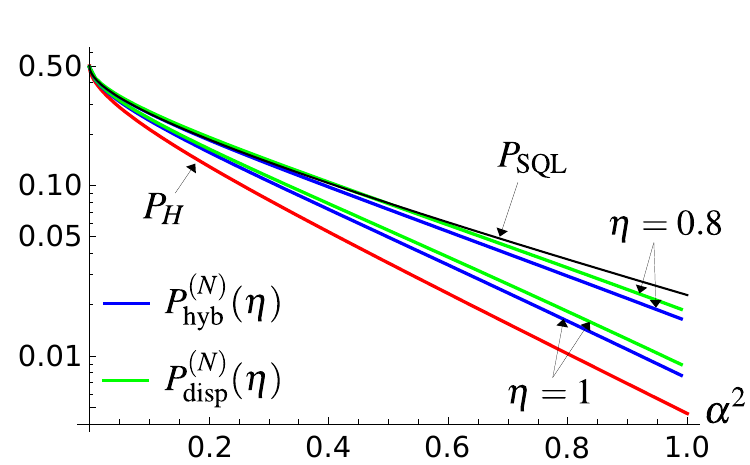}
\caption{Log plot of $\PHYB^{(N)}(\eta)$ and $\PDISP^{(N)}(\eta)$ as a function of the signal energy $\alpha^2$ for $N=1$ and different values of $\eta$. The PNR resolution is $M=2$.}\label{fig:06-ErrorP_eta}
\end{figure}

\begin{figure}[t]
\includegraphics[width=0.95\columnwidth]{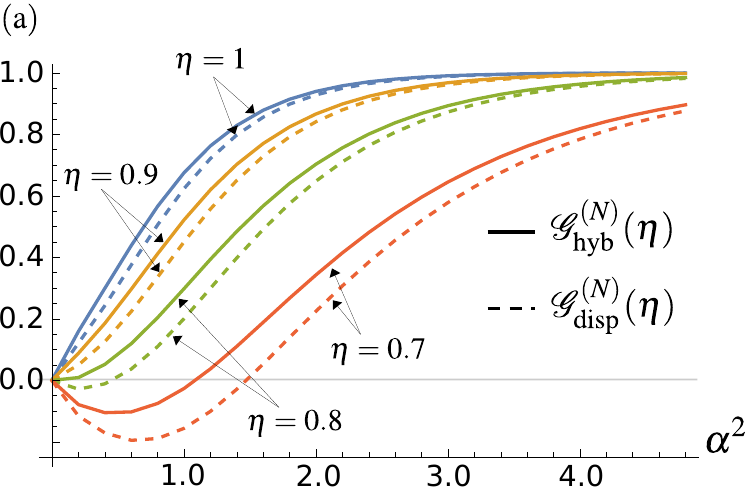}\\[4ex]
\includegraphics[width=0.95\columnwidth]{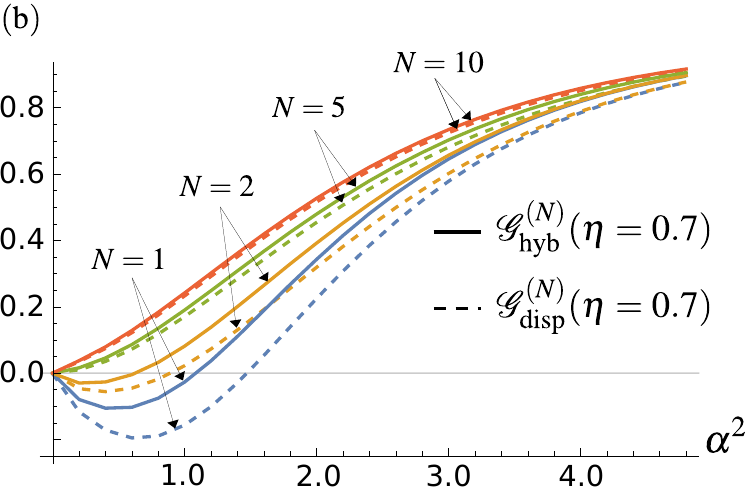}\\
\caption{(a) Plot of the gain ${\cal G}_{\p}^{(N)}(\eta) $, $\p=\disp,\hyb$, as a function of the signal energy $\alpha^2$ for $N=1$ and different quantum efficiency $\eta$. (Bottom) Plot of the gain ${\cal G}_{\p}^{(N)}(\eta) $, $\p=\disp,\hyb$, as a function of $\alpha^2$ for $\eta=0.7$ and different number of copies $N$. In both the pictures, the PNR resolution is $M=2$.}\label{fig:07-RatioSQL_eta}
\end{figure}

The error probability for the DFFRE $\PDISP^{(N)}(\eta)$ may be derived from the previous equations by fixing $\tau=1$. Plots of $\PHYB^{(N)}(\eta)$ and $\PDISP^{(N)}(\eta)$ are depicted in Fig.~\ref{fig:06-ErrorP_eta}, showing that the presence of a non-unit quantum efficiency increases the error probability, preventing the receivers to approach the Helstrom bound. Nevertheless, we still have $\PHYB^{(N)}(\eta) \le \PDISP^{(N)}(\eta)$ and, remarkably, in the high-energy regime both the discussed receivers beat the SQL~(\ref{eq:SQL}). To better highlight this feature, we consider the gain 
\begin{align}
{\cal G}_{\p}^{(N)}(\eta) = 1-\frac{P_{\p}^{(N)}(\eta)}{\PSQL} \, , \quad (\p=\disp,\hyb) \, ,
\end{align}
plotted in Fig.s~\ref{fig:07-RatioSQL_eta}(a) and (b). Accordingly, the SQL is outperformed when ${\cal G}_{\p}^{(N)}(\eta)\ge 0$.

If we consider a fixed number of copies $N$ [Fig.~\ref{fig:07-RatioSQL_eta}(a)], there exists a threshold energy $\alpha^2_{\p}(N,\eta)$ after which the discussed receivers beat the SQL, that is ${\cal G}_{\p}^{(N)}(\eta)\ge 0$ for $\alpha^2 \ge \alpha^2_{\p}(N,\eta)$. By reducing the quantum efficiency $\eta$, the gain and the threshold energy decrease and increase, respectively. More interestingly, in the opposite scenario where we fix $\eta$ and let $N$ vary [Fig.~\ref{fig:07-RatioSQL_eta}(b)], we see that increasing the number of copies mitigates the detriments of the quantum efficiency, and makes the gain increase. In particular, for a sufficiently large $N$, $\alpha^2_{\p}(N,\eta)$ may be made arbitrarily small, maintaining ${\cal G}_{\p}^{(N)}(\eta)\ge 1$ for all energies. In all cases, the HFFRE outperforms the DFFRE, as ${\cal G}_{\hyb}^{(N)}(\eta)\ge {\cal G}_{\disp}^{(N)}(\eta)$ and $\alpha^2_{\hyb}(N,\eta) \le \alpha^2_{\disp}(N,\eta)$.

\subsection{Dark counts $\boldsymbol\nu$}\label{sec:DC}

\begin{figure}[t]
\includegraphics[width=0.95\columnwidth]{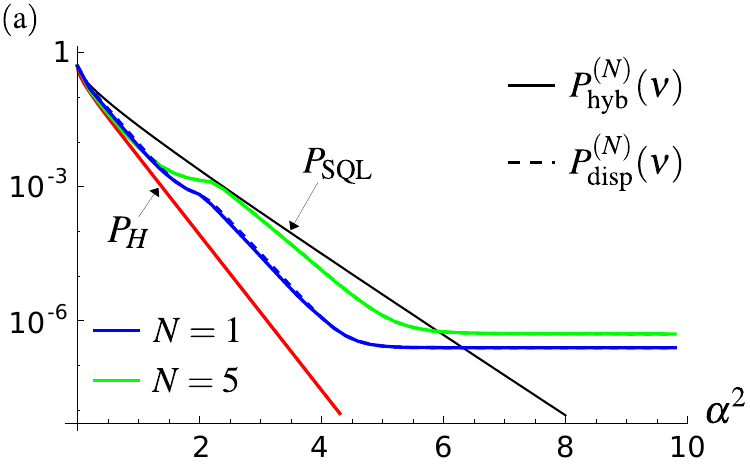}\\[2ex]
\includegraphics[width=0.95\columnwidth]{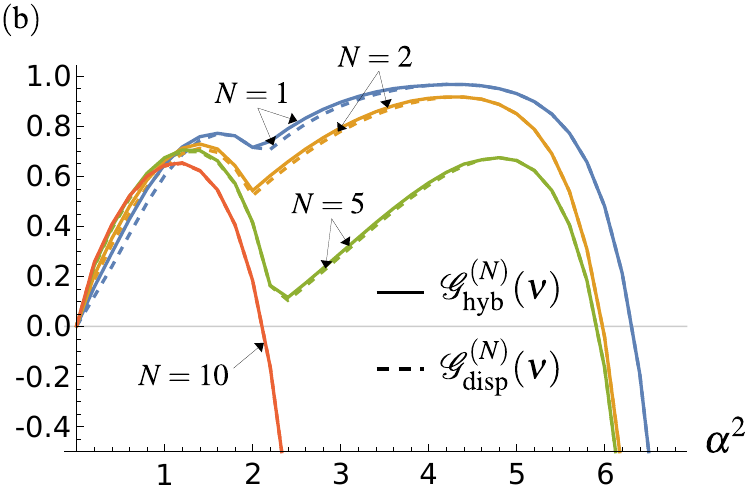}\\
\caption{(a) Log plot of $\PHYB^{(N)}(\nu)$ and $\PDISP^{(N)}(\nu)$ as a function of the signal energy $\alpha^2$ for different values of $N$. (b) Plot of the gain ${\cal G}_{\p}^{(N)}(\nu)$, $\p=\disp,\hyb$, as a function of $\alpha^2$ for different $N$. In both the pictures, the PNR resolution is $M=2$ and the dark count rate is $\nu=10^{-3}$.}\label{fig:08-DC}
\end{figure}

More drastic effects appear in the presence of dark counts.
With the term dark counts we refer to random clicks of the PNR$(M)$ detector induced by environmental noise, which may be modeled as Poisson events occurring at a given rate $\nu$ \cite{Humer2015}. The value of $\nu$ is strictly dependent on the response time of the particular detector employed, but in many cases we have $\nu \lesssim 10^{-3}$ \cite{Izumi2012, DiMario2018,Izumi2021,Thekkadath2021,Sidhu2021}. 
If the PNR$(M)$ statistics for an input coherent state $|\zeta\rangle$ follows the truncated Poisson distribution~(\ref{eq:PnPNRM}), in the presence of dark counts we still observe a truncated Poisson distribution, but with an increased rate $|\zeta|^2+\nu$ \cite{Notarnicola2023}.

Accordingly, in the presence of dark counts the HL probability distribution of the reflected signal $|\alpha^{(r)}_k\rangle$ reads
\begin{align}\label{eq: pDelta with DC}
{\cal S}_\Delta\Big(\nu;\alpha^{(r)}_k\Big) &=\sum_{n,m=0}^{M} p_n\biggl(\mu_{+}\Big(\alpha^{(r)}_k\Big)+\nu\biggr) \,
\notag\\[1ex]
& \hspace{1.0cm}
\times p_m\biggl(\mu_{-}\Big(\alpha^{(r)}_k\Big)+\nu\biggr) \, \delta_{(n-m), \Delta}
\end{align}
with the $\mu_{\pm}$ in Eq.~(\ref{eq:mupm}).
A more detrimental effect is observed in the displacement-photon counting scheme performed on the transmitted signal. Indeed, in the presence of dark counts the MAP criterion does not coincide anymore with on-off discrimination, as claimed in Sec.~\ref{sec: Disp}. 
On the contrary, in principle one should perform a different Bayesian inference process after each detection stage. However, for the sake simplicity here we adopt a simpler decision rule. We introduce a threshold outcome $1\le n_{\rm th}\le M$ such that if we get outcome $n<n_{\rm th}$ from the $(j-1)$-th PNR($M$) measurement we set $\sigma_j=\sigma_{j-1}$ and, then, displace the $j$-th copy by $D(\sigma_{j}\beta_j)$; otherwise if $n\ge n_{\rm th}$ we choose $\sigma_j=-\sigma_{j-1}$. In the ideal scenario of Sec.~\ref{sec: Disp} we have $n_{\rm th}=1$. The final decision rule becomes: $n<n_{\rm th} \rightarrow |-\sigma_{N} \alpha\rangle$ and $n\ge n_{\rm th} \rightarrow |\sigma_{N} \alpha\rangle$.

As a consequence, the correct decision probability ${\cal P}_{\rm hyb}^{(j)}(\nu;\tau,z)$ satisfies:
\begin{align}\label{eq:PcorrDC}
&{\cal P}_{\rm hyb}^{(j)}(\nu;\tau,z) =\notag\\[1ex]
&\hspace{0.5cm}\max_{\beta_{j}} \Bigg\{
{\cal P}_{\rm hyb}^{(j-1)}(\nu;\tau,z)\,
\widetilde{q}_0\big( \lambda_{-}^{(j)}(\sqrt{\tau}\alpha;\nu);n_{\rm th}\big)\notag\\[1ex]
&\hspace{1cm}+
\bigg[ 1 - {\cal P}_{\rm hyb}^{(j-1)}(\nu;\tau,z) \bigg]
\widetilde{q}_1\big(  \lambda_{+}^{(j)}(\sqrt{\tau}\alpha;\nu); n_{\rm th} \big)
\Bigg\} \, ,
\end{align}
where
\begin{align}\label{eq:q01TH}
\widetilde{q}_0(x;n_{\rm th}) &= \sum_{s=0}^{n_{\rm th}-1} e^{-x} \frac{x^s}{s!} \, ,  \\[2ex]
\widetilde{q}_1(x;n_{\rm th})  &= 1-\widetilde{q}_0(x;n_{\rm th})  \, ,
\end{align}
and
\begin{align}
\lambda_{\pm}^{(j)}(\alpha;\nu) = \lambda_{\pm}^{(j)}(\alpha)+\nu \, .
\end{align}
The initial condition of Eq.~(\ref{eq:PcorrDC}) reads
$${\cal P}_{\rm hyb}^{(0)}(\nu;\tau,z)=
\frac12 \Bigg[ \sum_{\Delta=-M}^{-1} {\cal S}_\Delta\Big(\nu;\alpha^{(r)}_1\Big)
+ \sum_{\Delta=0}^{M} {\cal S}_\Delta\Big(\nu;\alpha^{(r)}_0\Big) \Bigg] \, , $$ 
and the associated error probability is obtained as
\begin{align}\label{eq:Phyb_nu}
\PHYB^{(N)}(\nu) = 1- \max_{\tau,z, n_{\rm th}}{\cal P}_{\rm hyb}^{(N)}(\nu;\tau,z)  \, ,
\end{align}
where, differently from the other cases, we perform optimization also over the threshold discrimination outcome $n_{\rm th}$. As before, with the choice $\tau=1$ we retrieve the probability $\PDISP^{(N)}(\nu)$ associated with the displacement receiver.

The plots of $\PHYB^{(N)}(\nu)$ and $\PDISP^{(N)}(\nu)$ are reported in 
Fig.~\ref{fig:08-DC}(a) for different number of copies $N$ and $M=2$. 
The step-like behaviour of the curves follows from the adopted discrimination strategy: for $\alpha^2 \ll 1$ the optimized discrimination threshold is equal to $n_{\rm th}=1$, equivalent to on-off detection, whereas, for increasing $\alpha^2$, $n_{\rm th}$ jumps to higher integer values up to $n_{\rm th}=M$ in the regime $\alpha^2 \gg 1$. In turn, at every change in the threshold, the corresponding error probabilities exhibit a cusp.

Remarkably, in the presence of dark counts the performance of the receivers is not improving anymore with larger number of copies. In fact, increasing $N$ induces a reduction of the error probability only for $\alpha^2 \ll 1$. On the contrary, for large energies employing many copies becomes detrimental. Indeed, it has been shown in \cite{Notarnicola2023} that dark counts induce decision errors, letting the error probability saturate for $\alpha^2 \gg 1$. Accordingly, when we split the signal into $N$ copies, the decision errors induced by dark counts accumulate, letting the error probability reach higher saturating values. 

To quantify the present effect, some analytical results may be retrieved in the limit $\alpha^2 \gg 1$.
For the DFFRE, numerical results show that, in the regime $\alpha^2 \gg 1$, the optimized displacement amplitudes are $\beta_j\approx \alpha/\sqrt{N}$ and $n_{\rm th}=M$. Thus, we have $\lambda_{-}^{(j)}(\alpha;\nu)=\nu$ and $\lambda_{+}^{(j)}(\alpha;\nu)=\nu+ 4 \alpha^2/N \gg 1$. This implies that an error occurs only when the outcome $M$ is obtained from the input $|\alpha_0\rangle$, in turn the correct decision probability at the $j$-th step reads:
\begin{align}
{\cal P}_{\rm disp}^{(j)}(\nu) \approx {\cal P}_{\rm disp}^{(j-1)}(\nu) \, \widetilde{q}_0(\nu) + \left[1-{\cal P}_{\rm disp}^{(j-1)}(\nu) \right] \, .
\end{align}
By iteration, we get:
\begin{align}
\PDISP^{(N)}(\nu) \approx 1- \left\{\frac{\left[\widetilde{q}_0(\nu)-1\right]^N}{2} +\frac{1-\left[\widetilde{q}_0(\nu)-1\right]^N}{1-\left[\widetilde{q}_0(\nu)-1\right]} \right\} \, ,
\end{align}
being independent of the energy $\alpha^2$ and, therefore, letting the error probability saturate.
The same result also holds for the HFFRE, since the optimized transmissivity  $\tau_{\rm opt}$ in the high-energy regime is equal to $\tau_{\rm opt}=1$.

Finally, we note that the benefits of the hybrid scheme are more relevant for $N \lesssim 5$. For larger number of copies the improvement becomes negligible: as we can see, for instance, in Fig.~\ref{fig:08-DC}(a), the curves associated with the HFFRE and DFREE lines for $N=10$ are superimposed and fully indistinguishable.

The saturation of the error probability forbids to beat the SQL in the large energy regime. Indeed, the gain
\begin{align}
{\cal G}_{\p}^{(N)}(\nu) = 1-\frac{P_{\p}^{(N)}(\nu)}{\PSQL} \, , \quad (\p=\disp,\hyb) \, ,
\end{align}
plotted in Fig.~\ref{fig:08-DC}(b), is positive up to a maximum energy $\alpha^2_{\p}(N,\nu)$. Here the tradeoff between the number of copies and the error probability is clearer: for larger values of $N$ we increase ${\cal G}_{\p}^{(N)}(\nu)$ in the low-energy regime $\alpha^2 \ll 1$, at the expense of reducing also $\alpha^2_{\p}(N,\nu)$. If on the one hand we reduce the error probability for low energies, on the other one we inevitably reduce the range in which the receivers exhibit a quantum advantage.

\subsection{Visibility reduction $\boldsymbol\xi$}\label{sec:Vis}
\begin{figure}[t]
\includegraphics[width=0.95\columnwidth]{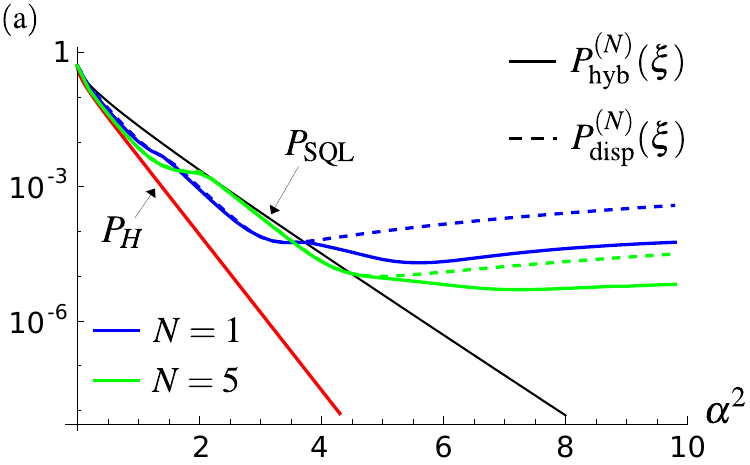}\\[2ex]
\includegraphics[width=0.95\columnwidth]{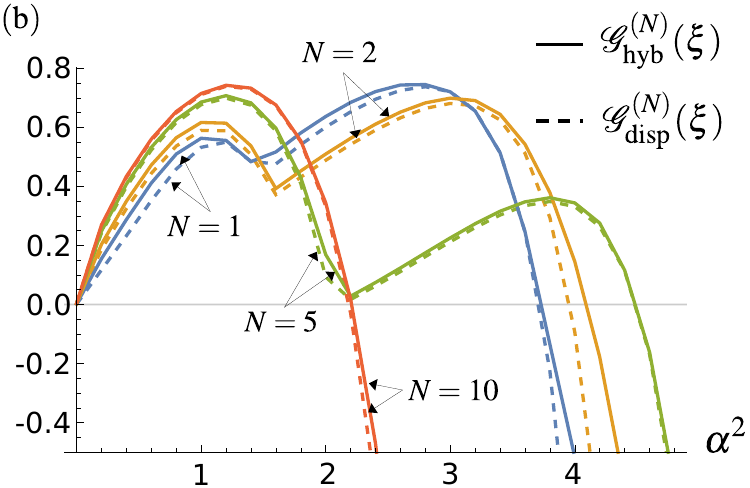}\\
\caption{(a) Log plot of $\PHYB^{(N)}(\xi)$ and $\PDISP^{(N)}(\xi)$ as a function of the signal energy $\alpha^2$ for different values of $N$. (b) Plot of the gain ${\cal G}_{\p}^{(N)}(\xi)$, $\p=\disp,\hyb$, as a function of $\alpha^2$ for different $N$. In both the pictures, the PNR resolution is $M=2$ and the visibility is $\xi=0.998$.}\label{fig:09-Vis}
\end{figure}

Visibility reduction is one of the typical effects associated with quantum interference. It is the direct consequence of the mode mismatch at the beam splitter which implements practically a displacement operation. The value $\xi\le1$ quantifies the spatial overlap between the signal field and the LO impinging at the beam splitter \cite{Becerra2015, DiMario2019}.

We model this imperfect mode matching in the following way. Suppose we want to displace a coherent state $|\zeta\rangle$ by a quantity $\beta$ to obtain $|\zeta+\beta\rangle$. For the sake of simplicity, we assume $\zeta,\beta \in \mathbb{R}$. Then the statistics of the subsequent PNR($M$) measurement will be a truncated Poisson distribution with rate $\mu= \zeta^2+\beta^2 + 2\xi \zeta \beta \neq (\zeta+\beta)^2$.

In the HFFRE we observe a visibility reduction both in the HL setup, where the signal is mixed with the LO $|z\rangle$, and in the conditional displacement operations governed by the feed-forward rule.
The HL probability distribution of the reflected signal $|\alpha^{(r)}_k\rangle$ becomes
\begin{align}\label{eq: pDelta with Vis}
{\cal S}_\Delta\Big(\xi;\alpha^{(r)}_k\Big) &=
\sum_{n,m=0}^{M} p_n\biggl(\mu_{+}(\alpha^{(r)}_k;\xi)\biggr)\notag\\[1ex]
&\hspace{1cm}\times p_m\biggl(\mu_{-}(\alpha^{(r)}_k;\xi)\biggr) \, \delta_{(n-m), \Delta}
\end{align}
with
\begin{align}\label{eq: mu with Vis}
    \mu_{\pm}(\alpha;\xi)&= \frac{\alpha^2 + z^2 \pm 2 \xi \ z \alpha}{2} \, .
\end{align}
For the feed-forward rule on the transmitted signal, we proceed as for the case of dark counts and introduce the threshold outcome $1\le n_{\rm th}\le M$.

Accordingly, the correct decision probability ${\cal P}_{\rm hyb}^{(j)}(\xi;\tau,z)$ satisfies:
\begin{align}\label{eq:PcorrVis}
&{\cal P}_{\rm hyb}^{(j)}(\xi;\tau,z) =\notag\\
&\hspace{0.5cm}\max_{\beta_{j}} \Bigg\{ {\cal P}_{\rm hyb}^{(j-1)}(\xi;\tau,z)\,
 \widetilde{q}_0\Big( \lambda_{-}^{(j)}(\sqrt{\tau}\alpha;\xi);n_{\rm th}\Big)\notag\\
&\hspace{1.0cm} + \bigg[ 1 -  {\cal P}_{\rm hyb}^{(j-1)}(\xi;\tau,z) \bigg]
 \widetilde{q}_1\Big(  \lambda_{+}^{(j)}(\sqrt{\tau}\alpha;\xi); n_{\rm th} \Big) 
 \Bigg\} \, ,
\end{align}
with the same $\widetilde{q}_{k}(x;n_{\rm th})$, $k=0,1$, introduced in Eq.~(\ref{eq:q01TH}), the rates
\begin{align}
\lambda_{\pm}^{(j)}(\alpha;\xi) = \frac{\alpha^2}{N} + \beta_j^2 \pm \frac{2 \xi \beta_j \alpha}{\sqrt{N}} \, ,
\end{align}
and the initial condition
$${\cal P}_{\rm hyb}^{(0)}(\xi;\tau,z)=
\frac12 \Bigg[ \sum_{\Delta=-M}^{-1} {\cal S}_\Delta\Big(\xi;\alpha^{(r)}_1\Big)
+ \sum_{\Delta=0}^{M} {\cal S}_\Delta\Big(\xi;\alpha^{(r)}_0\Big) \Bigg] \, . $$ 
Finally, the error probability writes:
\begin{align}\label{eq:Phyb_xi}
\PHYB^{(N)}(\xi) = 1- \max_{\tau,z, n_{\rm th}}{\cal P}_{\rm hyb}^{(N)}(\xi;\tau,z)  \, ,
\end{align}
whereas for $\tau=1$ we obtain the corresponding $\PDISP^{(N)}(\xi)$.

As depicted in Fig.~\ref{fig:09-Vis}(a), the behaviour of  $\PHYB^{(N)}(\xi)$ and $\PDISP^{(N)}(\xi)$ is similar to the case of dark counts, with a step-like behaviour induced by the jump in the threshold $n_{\rm th}$. Even in this case, increasing the number of copies $N$ reduces the error probability for low energies, $\alpha^2 \ll 1$, but, differently from Sec.~\ref{sec:DC}, this reduction holds also in the high-energy regime $\alpha^2 \gg 1$. In fact, the detriments of the visibility reduction are more relevant for strong signals and, in turn, the error probability for $\alpha^2 \gg 1$ becomes an increasing function of the energy \cite{Notarnicola2023}. As a consequence, splitting the incoming signal into a larger number of copies $N$ reduces the energy of each displaced copy, thus partially mitigating the effects of the imperfect displacements. 

With a similar argument to the one adopted for dark counts, we can obtain the analytic expression for the error probability in the high-energy regime. For the DFFRE, we have:
\begin{align}
\PDISP^{(N)}(\xi) \approx 1- \left\{\frac{\left[\widetilde{q}_0(g)-1\right]^N}{2} +\frac{1-\left[\widetilde{q}_0(g)-1\right]^N}{1-\left[\widetilde{q}_0(g)-1\right]} \right\} \, ,
\end{align}
with $g=2\alpha^2(1-\xi)/N$, being an increasing function of $\alpha^2$.
On the contrary, the HFFRE beats the DFFRE since the optimized transmissivity $\tau_{\rm opt}$ for the HFFRE is $<1$, and combining HL and displacement results in a lower error probability.

Anyway, there still exist an intermediate region, comprised between the regimes $\alpha^2 \ll 1$ and $\alpha^2 \gg 1$, where increasing $N$ is not beneficial anymore. Moreover, we note in the high-energy regime the HFFRE outperforms significantly the DFFRE, because of the higher degree of robustness of HL with respect to visibility reduction \cite{Notarnicola2023}.

The existence of three different energy regimes affects also the gain with respect to the SQL,
\begin{align}
{\cal G}_{\p}^{(N)}(\xi) = 1-\frac{P_{\p}^{(N)}(\xi)}{\PSQL} \, , \quad (\p=\disp,\hyb) \, ,
\end{align}
plotted in Fig.~\ref{fig:09-Vis}(b). As for the case of dark counts, we have ${\cal G}_{\p}^{(N)}(\xi)  \ge 0$ up to a maximum energy $\alpha^2_{\p}(N,\xi)$, but the behaviour of $\alpha^2_{\p}(N,\xi)$ is not monotonic with the number of copies $N$. For $N\lesssim 5$, splitting the signal into more copies improves the robustness of the quantum advantage, letting $\alpha^2_{\p}(N,\xi)$ increase. On the contrary, for larger $N$ the error probabilities surpass the SQL already in the intermediate energy regime and, in turn, $\alpha^2_{\p}(N,\xi)$ decreases.

\section{Conclusions}\label{sec: Concl}
In this paper we have proposed the HFFRE, a hybrid feed-forward receiver based on the combination of the HL setup introduced in \cite{Notarnicola2023}, the HYNORE, and the many-copy approach proposed in \cite{Sych2016}, that is the DFFRE. The key idea is to retrieve and exploit some a priori information on the phase of the incoming optical field to improve the sequence of conditional displacements of the DFFRE. Moreover, exploiting HL instead of standard homodyne detection allows to pursue this goal with sole PNR measurements.

We have investigated the performance of this hybrid scheme in comparison to the DFFRE and proved a reduction of the error probability in all conditions.
In particular, in ideal conditions the HFFRE is near-optimum and its error probability is closer to the Helstrom bound, asymptotically approaching the HYNORE \cite{Notarnicola2023}, whilst the DFFRE converges to the Kennedy receiver. 
In the presence of detection inefficiencies and visibility reduction the hybrid scheme turns out to be more robust with respect to the displacement one and still capable of outperforming the SQL in particular regimes.

The discussed hybrid method offers itself as a powerful approach to improve quantum receivers also for quaternary and $M$-ary PSK discrimination \cite{Izumi2012, Becerra2015, Izumi2013, Becerra2013, Izumi2020}, where displacement feed-forward schemes are less powerful and may benefit even more from a suitable combination with quadrature measurements.
Furthermore, the results obtained in the paper may foster further improvements in quantum communications as well as in quantum key distribution protocols \cite{Notarnicola2023-KB}.

\section*{Acknowledgements}
This work has been partially supported by the Ministry of Foreign Affairs and International Cooperation (MAECI), Project No.~PGR06314 ``ENYGMA''.

\appendix

\section{Derivation of the correct decision probability for the displacement feed-forward receiver}\label{app:derivation}
\begin{figure}[h]
\includegraphics[width=0.8\columnwidth]{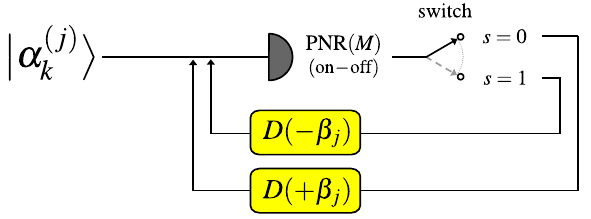}
\caption{Equivalent scheme of the DFFRE. Each copy $|\alpha_k^{(j)}\rangle$ undergoes a displacement operation whose sign is determined by the position of a switch $s$.}\label{fig:A-EqScheme}
\end{figure}

To derive the correct decision probability in Eq.~(\ref{eq:PcorrDISP}) we start noting that the DFFRE in Fig.~\ref{fig:02-DispScheme} is equivalent to scheme reported in Fig.~\ref{fig:A-EqScheme}.
It consists of a photon counter performing on-off detection connected to a switch $s$. After splitting the signal $|\alpha_k\rangle$ into the $N$ copies, we inject each copy $|\alpha_k^{(j)}\rangle$ one at a time into this equivalent setup.  The switch switches back and forth between two positions, called $s = 0$ and $s = 1$, with each click of the detector, applying alternatively two different optimized displacement operations
\begin{align}
&D(+\beta_j) \quad \mbox{if} \quad s^{(j)}=0 \, , \\
&D(-\beta_j) \quad \mbox{if} \quad s^{(j)}=1 \, .
\end{align}
In the above expression $s^{(j)}$ refers to the position of the switch after the $j$-th copy is processed.
The initial position of the switch is set to $s^{(0)}=0$, meaning that the first copy $|\alpha_k^{(1)}\rangle$ is displaced by $D(\beta_1)$, $\beta_1>0$. Then on-off detection is performed, implemented via a PNR$(M)$ measurement. After the first detection, if the detector clicks the position of the switch is changed to $s(1)=1$ and the second copy will be displaced by $D(-\beta_2)$,  $\beta_2>0$. Otherwise, we keep still $s(1)=0$ and the second copy will be displaced by $D(\beta_2)$. The feed-forward loop continues according to this basic rule: at every click of the on-off detector the switch changes its position. When all the $N$ copies are processed, the final decision is obtained by reading the position of the switch, according to:
\begin{align}
&s^{(N)}=0 \quad \rightarrow \quad \mbox{infer state } |\alpha_0\rangle \, , \\
&s^{(N)}=1 \quad \rightarrow \quad \mbox{infer state } |\alpha_1\rangle \, . 
\end{align}

We introduce the conditional probability $P_{kl}^{(j)}$ of inferring state ``$l$" after $j$ steps if signal ``$k$" is sent, $k,l=0,1$, $j=1,\ldots, N$. The correct decision probability then writes
$${\cal P}_{\disp}^{(j)} = \pi_0 P_{00}^{(j)} +\pi_1 P_{11}^{(j)}$$
with $\pi_0=\pi_1=1/2$.

At first we assume that state $|\alpha_0\rangle= |-\alpha\rangle$ is sent. Then, after $j$ steps a correct decision is performed in two cases: firstly if $s^{(j-1)}=0$ and the PNR($M$) detector does not click; secondly if $s^{(j-1)}=1$ and the PNR($M$) detector clicks. Accordingly we have:
\begin{align}\label{eq:P00}
P_{00}^{(j)}&=P_{00}^{(j-1)} q_{\rm off}\Big( \lambda_{-}^{(j)}(\alpha)\Big)
+ \bigg[1-P_{00}^{(j-1)}\bigg] q_{\rm on} \Big( \lambda_{+}^{(j)}(\alpha)\Big)
\end{align}
with the quantities introduced in Eq.s~(\ref{eq:q01}) and ~(\ref{eq:lambdapm}), to be solved with the initial condition
$$P_{00}^{(0)} =1 \, ,$$
as the switch is initialized in position ``0". In a similar way, we prove that $P_{11}^{(j)}$ satisfies the same equation with initial condition $P_{11}^{(0)}=0$. We conclude that also ${\cal P}_{\disp}^{(j)}$ is a solution of Eq.~(\ref{eq:P00}) with ${\cal P}_{\disp}^{(0)}=1/2$. Moreover, the displacement amplitudes $\beta_j$ may be optimized to maximize ${\cal P}_{\disp}^{(j)}$ in each step.

\newpage

\end{document}